\documentclass{article}

\usepackage{citesort}
\def\ud{\mathrm{d}}
\def\si{\sigma}
\def\de{\delta}
\def\De{\Delta}
\def\na{\nabla}

\def\pa{\partial}
\def\fr{\frac}

\begin{document}
\vspace*{1.0cm}
\noindent
{\bf
{\large
\begin{center}
Comments on some recently proposed experiments that should distinguish Bohmian mechanics from quantum mechanics
\end{center}
}
}

\vspace*{.5cm}
\begin{center}
W.\ Struyve, W.\ De Baere
\end{center}

\begin{center}
Laboratory for Theoretical Physics\\
Unit for Subatomic and Radiation Physics\\
Proeftuinstraat 86, B--9000 Ghent, Belgium\\
E--mail: willy.debaere@rug.ac.be, ward@odiel.rug.ac.be
\end{center}

\begin{abstract}
\noindent
Recently Ghose \cite{ghose1,ghose2,ghose4} and Golshani and Akhavan \cite{golshani1,golshani2,golshani3} claimed to have found experiments that should be able to distinguish between Standard Quantum Mechanics and Bohmian Mechanics. It is our aim to show that the claims made by Ghose, Golshani and Akhavan are unfounded. 
\end{abstract}

\section{Introduction} 

According to Standard Quantum Mechanics (SQM), the complete description of a physical system is provided by its wave function. In Bohmian Mechanics (BM){\footnote{For a mathematical review see \cite{holland}.}} the standard description of quantum phenomena, by means of the wavefunction $\psi$, is enlarged by considering particles that follow definite tracks in space-time (dependent on the initial conditions). These positions of a particle on these tracks act as the hidden variables of SQM. The positions of the particles are hidden because BM is constructed in a way to give the same statistical predictions as SQM if a measurement is performed. This is accomplished by assuming the probability distribution for an ensemble in  BM to be the same as the quantum mechanical distribution. This distribution is called the {\em quantum equilibrium distribution} (see section 2).

Yet, recently Ghose \cite{ghose1,ghose2,ghose4} and Golshani and Akhavan \cite{golshani1,golshani2,golshani3} (short GGA) proposed some experiments that should be able to distinguish between SQM and BM at the level of individual detections. It is the aim of the present work, however, to show that the claims made by GGA are unfounded. Moreover it should be clear that, with the {\em quantum equilibrium hypothesis} in mind, one cannot obtain a distinguishment between SQM and BM. Only a modified or extended Bohmian theory can yield experimentally observable differences with SQM. This stresses once more the fact that BM is nothing more than a possible (causal) interpretation of SQM, as is clearly stated already in the beginning by Bohm \cite{bohm1,bohm2}.   

\section{Quantum equilibrium hypothesis}

In SQM a physical system is described in configuration space by its wavefunction $\psi({\bf x}_1,\dots,{\bf x}_n,t)$, dependent on $n$ 3-vectors ${\bf x}_j$. This wavefunction obeys the Schr\"odinger equation
\begin{equation}
i \hbar \frac{\partial \psi({\bf x}_1, \dots,{\bf x}_n,t)}{\pa t} = \hat{H} \psi({\bf x}_1, \dots,{\bf x}_n,t)
\label{12}
\end{equation}
Given an initial wavefunction $\psi({\bf x}_1,\dots,{\bf x}_n,0)$ this  equation can be solved to give a unique solution $\psi({\bf x}_1,\dots,{\bf x}_n,t)$. When a position measurement is performed on an ensemble of identically prepared systems (all described by the same wavefunction), the probability $P({\bf Q} _1 ,\dots,{\bf Q}_n,t_0)$ of making a joint detection at a certain time $t_0$ of the $n$-particles at positions ${\bf Q} _1 ,\dots,{\bf Q}_n$ in physical space is given by
\begin{equation}
P({\bf Q} _1 ,\dots,{\bf Q}_n,t_0) = \psi^* ({\bf Q}_1, \dots ,{\bf Q}_n,t_0)  \psi ({\bf Q}_1, \dots ,{\bf Q}_n,t_0)
\label{14}
\end{equation}

In BM, SQM is considered as an incomplete theory. Apart from a wavefunction (obeying (\ref{12})) one introduces additional (hidden) variables to describe the physical system. These hidden variables are $n$ vectors that have to be interpreted as {\em actual} position vectors ${\bf X}_k(t)$ associated with $n$ particles in 3-dimensional physical space. According to BM these vectors are also the position vectors revealed in a position measurement. This is contrary to SQM where no particles exist as localized entities, i.e.\ as entities that have position vectors, until a position measurement is performed.

Bohm \cite{bohm1,bohm2} obtained the laws of motion for the particles by giving a new interpretation to the real and imaginary part of the Schr\"odinger equation. The real part is interpreted as a classical Hamilton-Jacobi equation with an additional quantum mechanical potential, the {\em quantum potential}. This interpretation leads to the following differential equations for the position vectors ${\bf X}_k(t)$ 
\begin{equation}
\frac{\ud {\bf X}_k}{\ud t} = \frac{\hbar}{m_k} \textrm{Im} \fr{\psi^* ({\bf x}_1, \dots ,{\bf x}_n,t) {\bf{\na}}_k \psi ({\bf x}_1, \dots ,{\bf x}_n,t)}{|\psi({\bf x}_1, \dots ,{\bf x}_n,t)|^2} \Big|_{{\bf x}_j={\bf X}_j}
\label{13}
\end{equation}
where $m_k$ is the mass of the $k^{th}$ particle. Once we have a solution for equation (\ref{12}), equation (\ref{13}) can be solved given the initial positions ${\bf X}_k(0)$. In this way the $n$ actual position vectors  ${\bf X}_k(t)$  of the particles are uniquely determined. If we then consider an ensemble of systems, all described by the same wavefunction, then this ensemble determines a probability distribution $\rho({\bf X}_1, \dots,{\bf X}_n,t)$ of the actual position vectors of the $n$ particles. This is the distribution that would be obtained, according to BM, when a position measurement on an ensemble were performed. If we want BM to give the same predictions as SQM in a position measurement, then the probability distribution $P$ of SQM in equation (\ref{14}), has to be the same as the probability distribution $\rho$ of BM, i.e. we must have
\begin{equation}
\rho({\bf X}_1, \dots,{\bf X}_n,t) = |\psi ({\bf X}_1, \dots,{\bf X}_n,t)|^2 
\label{15}
\end{equation}
for all times $t$. If this equality is assumed, and this is what is done in BM \cite{bohm1,bohm2}, the imaginary part of the Schr\"odinger equation 
\begin{equation}
\fr{\pa |\psi|^2}{\pa t} + \sum_k {\bf{\na}}_k \cdot ({\bf{v}}_k |\psi|^2) =0
\label{16.1}
\end{equation} 
with
\begin{equation}
{\bf{v}}_k = \frac{\hbar}{m_k} \textrm{Im} \fr{\psi^*  {\bf{\na}}_k \psi }{|\psi|^2} 
\label{16.2}
\end{equation} 
is the continuity equation, describing the conservation of the probability density of the particles. In fact it is sufficient that we assume
\begin{equation}
\rho({\bf X}_1, \dots,{\bf X}_n,t_0) = |\psi ({\bf X}_1, \dots,{\bf X}_n,t_0)|^2 
\label{16}
\end{equation}
at a certain time $t_0$ (for example at $t=0$) because both $|\psi|^2$ and $\rho$ satisfy the continuity equation. Thus, as far as predictions involving particle positions are concerned, BM is in complete accordance with SQM if the initial particle positions ${\bf X}_k(0)$ are distributed according to $|\psi({\bf X}_1, \dots,{\bf X}_n,0)|^2$ in the ensemble. This is what is called the {\em quantum equilibrium hypothesis} (QEH) by D\"urr, Goldstein and Zangh\`\i   \cite{durr} and the distibution is called the {\em quantum equilibrium distribution}. Because every measurement is in fact a position measurement, it is clear that there can {\bf never} be an experimental difference between BM and SQM, as a result of the QEH. This implies also that, despite of definite trajectories in BM, we can only predict and verify relative frequencies. So an individual event cannot be studied independently from the ensemble. 

The only difference that remains between BM and SQM is an interpretational one. In BM $\rho ({\bf Q} _1 ,\dots,{\bf Q}_n,t_0)$ is interpreted as the probability of the particles really {\em being} at the positions ${\bf Q} _1 ,\dots,{\bf Q}_n$ at time $t_0$ whereas in SQM $P({\bf Q} _1 ,\dots,{\bf Q}_n,t_0)$ is the probability of the particles {\em being detected} at the positions ${\bf Q} _1 ,\dots,{\bf Q}_n$ at time $t_0$.

The reason why we cannot distinguish the two theories is that we cannot observe a particle without disturbing its movement. I.e.\ we cannot obtain knowledge of the position of the particle without changing its wavefunction (this is the collapse in SQM). This changing wavefunction leads then to changing particle velocities (as follows from ({\ref{13}})), leaving a disturbed system. The best example of this is the diffraction at a slit: the smaller the slit (i.e.\ the better we try to get the initial position in the slit), the wider the scattering angle. 
In this way a quantum mechanical measurement is very different from a measurement in classical mechanics, where trajectories of objects are generally accepted because one can infer successive positions of an object without disturbing its motion, for example by using light that scatters from the object.

We want to remark that the QEH was already postulated by Bohm in order to assure complete equivalence between BM and SQM. So BM does not provide us with new experimentally verifiable predictions, but instead gives us a broader conceptual framework that may serve as a basis for new or modified mathema\-tical formulations for the description of physical systems. In such theories the QEH will evidently break down and this is what Bohm meant in {\cite{bohm2}}:
\begin{quote}
``An experimental choice between these two interpretations cannot be made in a domain in which the present mathematical formulation of the quantum theory is a good approximation, but such a choice is conceivable in domains, such as those associated with dimensions of the order of $10^{-13}$ cm, where the extrapolation of the present theory seems to break down and where our suggested new interpretation can lead to completely different kinds of predictions.''
\end{quote}  
Such modifications and extensions of BM are for example given by Bohm himself in \cite{bohm1,bohm3,bohm4,bohm5}.

\section{Outline and discussion of the experiments}

The proposed experiments of Ghose \cite{ghose1,ghose2} and Golshani and Akhavan \cite{golshani1,golshani2} make use of a pair of identical, non-relativistic, bosonic particles labeled 1 and 2. The particles emerge pair by pair (so there is only one pair in the device at a time) from a  source placed in front of a screen with two identical slits A and B, with coordinates $( 0, \pm Y)$. The particles are simultaneously diffracted by the two slits.  We can suppose the wavefunction of the system, in the $x-y$ plane, after diffraction, to be of the form 
\begin{equation}
\psi(x_1,y_1,x_2,y_2,t) = N[\psi_A(x_1,y_1,t) \psi_B(x_2,y_2,t) +\psi_B(x_1,y_1,t) \psi_A(x_2,y_2,t) ]
\label{1}
\end{equation}
where $\psi_A$ is the diffracted wave coming from slit A, and $\psi_B$ is the one coming from slit B. This wavefunction can be considered to describe particles going to different slits. The case in which a pair of particles can  ``travel"  through one slit at a time (\cite{golshani1,golshani2}) is not considered here. The two particles are then simultaneously detected at a fixed screen parallel with the $y$-axis. If we suppose the detectors to be idealized pointdetectors, then SQM gives the following probability for detecting the pair of particles (in the ensemble) at time $t_0$ at positions $y_1 = Q_1$ and $y_2 = Q_2$ on the screen:
\begin{equation}
P(Q_1,Q_2,t_0)= | \psi(y_1,y_2,t) |^2 \big|_{y_1 = Q_1,y_2 = Q_2,t=t_0} =  | \psi(Q_1, Q_2,t_0) |^2
\label{4}
\end{equation} 
If the detectors detect over regions $\De Q_1$ and  $\De Q_2$ then the probability of joint detection is given by
\begin{equation}
\bar{P}_{12}= \int^{Q_1 + \De Q_1}_{Q_1} \int^{Q_2 + \De Q_2}_{Q_2} \ud y_1 \ud y_2  | \psi(y_1,y_2,t) |^2
\label{46}
\end{equation} 
According to the QEH these probabilities must be the same for BM. 

We will now first look at Ghose's explanation of how a distinguishment can be achieved \cite{ghose1,ghose2}. Using plane waves for $\psi_A$ and $\psi_B$ \cite{ghose2} or more general using the natural symmetry of the setup (the slits are symmetrically placed around the $x$-axis) together with the bosonic symmetry \cite{ghose1}, Ghose finds that the $y$-coordinates of the Bohmian trajectories (calculated with ({\ref{13}})) satisfy{\footnote{Similar equations are obtained with spherical waves for $\psi_A$ and $\psi_B$ \cite{marchildon1,ghose3} or Gaussian waves (see below).}} 
\begin{equation}
{\dot y}_1  + {\dot y }_2 =0  \qquad \textrm{or}  \qquad y_1 (t) + y_2 (t) = y_1 (0) + y_2 (0)
\label{1.1}
\end{equation}
If the two particles depart initially symmetrically about the $x$-axis, i.e.
\begin{equation}
y_1 (0) + y_2 (0) = 0
\label{1.2}
\end{equation}
 then the particles will remain symmetrically about the $x$-axis for all times. This leads Ghose to state the following{\footnote{Taken from {\cite{ghose2}} adjusted with the factor $1/N$ and adapted to our notations.}
\begin{quote}
``\dots consider then the Bohm ensemble to be built up of single pairs of particle trajectories arriving at the screen at different instants of time $t_i$ such that the joint probability of detection is given by
\begin{eqnarray}
P^*_{12} &=& \lim_{N \to \infty} \sum^N_{i=1} \frac{1}{N \de(0)} \int \ud y_1 \ud y_2 P(y_1,y_2,t_i) \nonumber \\
&& \times \de (y_1 - y_1(t_i)) \de (y_2 - y_2(t_i)) \de (y_1(t_i) + y_2(t_i))
\nonumber \\
&=& \lim_{N \to \infty} \sum^N_{i=1} \frac{1}{N} P(y_1(t_i),-y_1(t_i),t_i) = 1  
\label{8}
\end{eqnarray}
where the constraint ({\ref{1.2}}) has been taken into account. Every term in the sum represents only one pair of trajectories arriving at the screen at the points $(y_1(t_i),-y_1(t_i))$ at time $t_i$ \dots''
\end{quote}
Ghose then concludes that when the detectors are placed symmetrically about the $x$-axis, they will record coincidence counts, just as SQM predicts. But if they are placed asymmetrically, the joint detection of every pair, and hence also their time average, will produce a null result which is not predicted by SQM. Instead SQM 
gives the probability $\bar{P}_{12}$ (in equation (\ref{46})) for joint detection.  This would be the incompatibility between SQM and BM. However, it is clear that $P(y_1(t_i),-y_1(t_i),t_i) \ne 1$ for every chosen couple $(y_1(t_i),-y_1(t_i))$, because of the identity $P \equiv |\psi|^2$. In addition, it is impossible to fix the intitial positions of every pair in order to have $y_1 (0) + y_2 (0) = 0$, without making measurements. We will discuss this in depth by using Gaussian waves. Following thereby Golshani and Akhavan \cite{golshani1,golshani2}, the wavefunction in the $x$-direction is a plane wave and in the $y$-direction the slits generate Gaussian profiles
\begin{eqnarray} 
\psi_{A,B}(x,y,t) &=& (2\pi \si^2_t)^{-1/4} e^{-(\pm y-Y - u_y t)^2/4\si_0 \si_t + i[ k_y(\pm y-Y - u_y t/2) ]} \nonumber\\ 
&& \times e^{i[k_x x-Et/\hbar]} 
\label{2}
\end{eqnarray}
where
\begin{eqnarray}
\si_t &=& \si_0 (1 + \frac{i\hbar t}{2m\si^2_0}) \\
u_{x,y} &=& \frac{\hbar k_{x,y}}{m}\\
E &=& \frac{1}{2} m u^2_x
\label{3} 
\end{eqnarray}
The motion in the $x$-direction is irrelevant and will be suppressed. If one defines $y = (y_1 + y_2)/2$ as the centre of mass in the $y$-direction, then one finds by using (\ref{13}), (\ref{1}) and (\ref{2}) that{\footnote {Taken from \cite{golshani1,golshani2}.}
\begin{equation}
\dot{y} = \frac{(\hbar/2m\si^2_0)^2}{1+(\hbar/2m\si^2_0)^2 t^2} yt
\label{6}
\end{equation}
which yields after integration
\begin{equation}
y(t) = y(0) \sqrt{1+(\hbar/2m\si^2_0)^2 t^2}
\label{7}
\end{equation}
If at $t=0$ the centre of mass of the particles is exactly on the $x$-axis (i.e. $y(0)=0$) then the centre of mass will remain on the $x$-axis for all time. So according to BM the particles will always be detected symmetrically about the $x$-axis if $y(0)=0$ for each pair of particles. Then Golshani and Akhavan argue in the line of Ghose that this property leads to a distinguishment when the time ensemble ({\ref 8}) is considered.
 
There are several remarks in order:
\begin{enumerate}
\item Although the probability of detecting every pair of particles symmetrically about the $x$-axis is equal to one if equation (\ref{1.2}) is assumed valid, $P(y_1(t_i),-y_1(t_i),t_i)$ cannot be equal to 1 for every pair $(y_1(t_i),-y_1(t_i))$ at time $t_i$, because $P(y_1,y_2,t) = |\psi(y_1,y_2,t) |^2$ (together with (\ref{1})). So $P^*_{12}$ cannot be equal to 1 .   
\item We can also not restrict the particles to depart initially symmetrically about the $x$-axis without changing the wavefunction, because $|\psi|^2$ does not contain the constraint $y(0)=0$, or put in another way $\psi$ does not restrict $y(0)$ from being different from zero. The initial positions of the particles (in BM) are distributed according to the absolute square of the wavefunction. So the initial coordinate of the centre of mass $y(0)$ is in the same way distributed. Given a system with wavefunction $\psi$ then we cannot determine the initial conditions better than provided by the wavefunction $\psi$ (i.e.\ we may not violate the QEH). Thus if we want $y(0)$ to be zero then we have to change the wavefunction $\psi$ by making measurements.

The fact that the initial positions have to be distributed according to the absolute square of the wavefunction has also been noted by Marchildon \cite{marchildon2}, who goes on by demonstrating in specific cases that BM and SQM indeed lead to the same predictions if no restriction on $y(0)$ is supposed. 
\item As an example we will show what happens if we take the slits very narrow in order to make sure that the particles depart symmetrically. Small slits imply a small $\si_0$ in (\ref{2}). Looking at (\ref{7}) one sees that $y(t)$ becomes large, although $y(0)$ is small, if time increases. This leads to a high probability of assymmetrical detection on the screen. In order to assure that $y(t) \approx y(0) $ one could propose to make $m$ large. So the particles would remain symmetrically in BM. But then $\psi$ in equation (\ref{1}) contains sharply peaked gaussians that do not spread rapidly, because then $u_y \ll$ and $\si_t \ll$. So also SQM then predicts a symmetrical, joint detection.    
\item In addition to the described experiment, Golshani and Akhavan \cite{golshani1,golshani2} proposed another one which uses selective detection. Suppose that $y(0)= \de$ and that each detector is placed on one side of the $x$-axis. If one detector is placed approximately on the $x$-axis then, from equation (\ref{7}), the other detector will never detect the other particle within a distance $L \ge \de$. If every pair in the ensemble is restricted in this way, there would be an observable difference. But because Golshani and Akhavan again have to suppose restricted initial positions, they again violate the QEH.

They even claim \cite{golshani2} that one can make $\De y (0)  \ll \si_0$, despite of the QEH. But little calculation shows that if the overlap of the wavefuncion is negligible at $t=0$, i.e.\ $\si_0$ is small compared to $Y$ (so crossterms in expectation values can be dropped), then $\De y(0) = \frac{1}{\sqrt{2}} \si_0$. One cannot adjust $y_0$ independently of $\si_0$, again as a consequence of the QEH.
\item In \cite{golshani3} Golshani and Akhavan propose an experiment similar to the one described above. Because the distinguishment between the two theories is based on the same grounds, the same arguments apply to that experiment as well.
\item 
If it would be the purpose of GGA to contest the QEH, then they should have to propose an alternative mathematical formulation of BM. But this is not what is suggested by GGA in \cite{ghose1,ghose2,ghose4,golshani1,golshani2,golshani3} and \cite{ghose5}. By the way, contesting the QEH could be done with less complicated experiments.
\end{enumerate}
 
\section{Ergodicity}
According to Ghose \cite{ghose4,ghose3} the incompatibity between BM and SQM in the two-slit experiment described above arises from the non-ergodic properties of the Bohmian description of the system, whereas SQM would be ergodic for every system. This was the reply of Ghose to the paper of Marchildon \cite{marchildon2}. We will now show that Ghose's definitions of ergodicity for SQM can be adopted to BM as well and on the other hand that his proof of the non-ergodicity of BM can be refuted. For definitions and theorems concerning ergodicity we refer to the book of Arnold and Avez.

Let us first look at the proof of Ghose \cite{ghose4} for the ergodicity of SQM of an arbitrary two particle system with wavefunction $\psi(x_1,x_2,t)$. We can write $\psi$ in the basis of orthonormal energy eigenfunctions $\phi_n(x_1,x_2)$
\begin{equation}
\psi(x_1,x_2,t) = \sum_n c_n e^{-iE_nt/\hbar} \phi_n(x_1,x_2)
\label{}
\end{equation} 
With the time and space average  for any observable $\hat{F}$, denoted respectively $F^*$ and $\bar{F}$, Ghose proves that   
\begin{eqnarray}
F^* &=& \lim_{T \to \infty} \frac{1}{T} \int^T_0 \ud t \int \ud x_1 \ud x_2 \psi^*(x_1,x_2,t) \hat{F}  \psi(x_1,x_2,t) \nonumber\\
&=& \sum_n |c_n|^2 \int \ud x_1 \ud x_2 \phi^*_n(x_1,x_2) \hat{F}  \phi_n(x_1,x_2) \nonumber\\
&=& \bar F
\label{17}
\end{eqnarray} 
This defines $F^*$ and $\bar{F}$. The equality of space and time averages for an arbitrary observable $\hat{F}$ induces ergodicity for SQM. Note that $\bar F$ is not the entire space average of $\hat F$ at $t=0$, but only an average over the diagonal elements. So, regardless the question wether or not it is usefull to introduce ergodicity in this way for quantum mechanics, the proof above can be adopted without problems to the case of BM as well. This is done by noting that the {\em complete} space average of the observable $\hat F$
\begin{equation}
\int \ud x_1 \ud x_2 \psi^*(x_1,x_2,t) \hat{F}  \psi(x_1,x_2,t)
\end{equation}
in (\ref{17}) equals the space average in BM of the ``local expectation value'' $F(x_1,x_2,t)$ of the hermitian  operator ${\hat F}$ defined by
\begin{equation}
F(x_1,x_2,t) = \textrm{Re}  \frac{\psi^*(x_1,x_2,t) \hat{F}  \psi(x_1,x_2,t) } {\psi^*(x_1,x_2,t)  \psi(x_1,x_2,t)}
\label{18}
\end{equation}
Hence $F$ can be interpreted as a property of the particle in BM (see \cite{holland}). The space average in BM is then
\begin{eqnarray}
&&\int  \ud x_1 \ud x_2  P(x_1,x_2,t) F(x_1,x_2,t) \nonumber\\
&&=  \textrm{Re}\int  \ud x_1 \ud x_2  \psi^*(x_1,x_2,t) \hat{F}  \psi(x_1,x_2,t) \nonumber\\
&&=  \int  \ud x_1 \ud x_2  \psi^*(x_1,x_2,t) \hat{F}  \psi(x_1,x_2,t)
\label{44}
\end{eqnarray}
where the hermiticity of ${\hat F}$ has been taken into account. So, the space average in BM equals the space average in SQM. This introduces ergodicity, as introduced by Ghose, for BM as well.

The non-ergodicity of BM was proved in \cite{ghose5}, by showing that the 2-slit system is decomposable, thereby again assuming the condition ({\ref{1.2}}). But relinquishing this condition can lead to ergodicity of BM. Non-ergodicity of the Bohmian system would imply that the space average $\bar{P}_{12}$ (from (\ref{46})) and the time average $P^*_{12}$ (from (\ref{8})) are not equal. The explanation of Ghose is that
\begin{equation}
P^*_{12} = 0 \quad \textrm{and} \quad \bar{P}_{12} \ne 0
\label{47}
\end{equation}
if the detectors are placed sufficiently asymmetrical. Apart from the discussion on the validity of expression (\ref{8}), already held in the previous section, we want to quote a theorem of G.D.\ Birkhoff and A.J.\ Khinchin \cite{arnold} which states that for a dynamical system $(M,\mu,\phi_t)$ and $f \in L_1(M,\mu)$ a complex valued $\mu$-summable function on $M$, the following equality holds
 \begin{equation}
\int_M  f^* \ud \mu = \bar{f}
\label{48}  
\end{equation}
This is clearly in contradiction with (\ref{47}).

In \cite{ghose4} Ghose provides us with an additional example of a non-ergodic system in BM. The system is the quantum mechanical analog of two coupled harmonic oscillators. Ghose comes to the same conclusions as for the 2-slit systems, using the same arguments. So my counter-arguments apply as well.

\section{Conclusion}
The cause of the different predictions made by BM and SQM, is that both theories different wavefunctions were used for the calculations of the probability distributions. BM used a wavefunction with restricted initial positions while SQM did not. Thus a new wavefunction has to be introduced, if we want to restrict the intitial conditions, which leads again to the same predictions as SQM. The trajectories that are hidden in BM cannot be revealed without disturbing the system.

\vspace*{6mm}
\noindent
Acknowledgements:

\vspace*{6mm}
\noindent
WS acknowledges financial support from the F.W.O.\ Belgium.


\end{document}